\documentclass[twocolumn,showpacs,preprintnumbers,amssymb]{revtex4}

\usepackage{graphicx}
\usepackage{epsfig}
\usepackage{dcolumn}
\usepackage{bm}

\def\gsim{\lower0.5ex\hbox{$\:\buildrel >\over\sim\:$}}
\def\lsim{\lower0.5ex\hbox{$\:\buildrel <\over\sim\:$}}

\newcommand{\spur}[1]{\not\! #1 \,}

\begin{document}

\title{
$B\to \phi \pi$ and $B^0 \to \phi\phi$ in the Standard Model
and new bounds on R parity violation
}

\author{Shaouly Bar-Shalom}
\affiliation{Theoretical Physics Group, Rafael, Haifa 31021, Israel}
\author{Gad Eilam}
\author{Ya-Dong Yang}
\affiliation{Physics Department, Technion-Israel Institute of
Technology, Haifa
32000, Israel}
%

\begin{abstract}
We study the pure penguin decays $B \to \phi\pi$ and $B^0 \to \phi\phi$. 
Using QCD factorization, we find  
${\cal B}(B^\pm \to\phi\pi^{\pm} )=2.0^{+0.3}_{-0.1}\times 10^{-8}$. 
For the  pure penguin annihilation process $B^0 \to \phi\phi$,
analyzed here for the first time, 
${\cal B}(B^0 \to\phi\phi)=2.1^{+1.6}_{-0.3}\times 10^{-9}$. 
The smallness of these decays in the Standard Model makes them 
sensitive probes for new physics.\ From the upper limit of $B\to
\phi\pi$,
we find constraints on R parity violating couplings,
$| \lambda{''}_{i23}\lambda{''}_{i21}|<6\times10^{-5}$, 
$| \lambda'_{i23}\lambda'_{i21}|<4\times10^{-4}$ and  
$| \lambda'_{i32}\lambda'_{i12}|<4\times10^{-4}$ for $i=1,2,3$.
Our new bounds on $|\lambda{''}_{i23}\lambda{''}_{i21}|$ are one order
of magnitude stronger than before.
Within the available upper bounds for 
$| \lambda{''}_{i23}\lambda{''}_{i21}|$,
$|\lambda'_{i23}\lambda'_{i21}|$ 
and $|\lambda'_{i32}\lambda'_{i12}|$, we find that 
${\cal B}(B\to\phi\phi)$ 
could be enhanced to $10^{-8}\sim 10^{-7}$. Experimental searches for
these
decays are strongly urged.
\end{abstract}



\maketitle

Charmless two-body nonleptonic decays of 
$B$ mesons provide tests for the Standard Model (SM) at both tree
and loop levels. They also test hadronic 
physics and probe  possible flavor physics beyond the SM.  
In past years, we have witnessed considerable progress in
studies of these decays.  Many such processes
have been measured or upper-limited  \cite{pdg2000}. Theoretically, 
QCD factorization in which non-factorizable effects
are calculable was presented  \cite{BBNS},
while there is also progress in perturbative QCD
approaches  \cite{lisanda}. 

In this letter,  we will use QCD factorization 
to study $B^{0,\pm}\to \phi\pi^{0,\pm}$ and $B^0 \to \phi\phi$.
In  \cite{fleischer,DU}, 
$B^{0,\pm}\to \phi\pi^{0,\pm}$, dominated by
electroweak penguins, were studied by employing naive factorization. 
By naive factorization one means that the hadronic matrix elements of
the
relevant four-quark operators are factorized into the product of
hadronic 
matrix elements of two quark currents that are described by form factors 
and decay constants.  In contrast to  color allowed processes, where
naive factorization works reasonably
well, this assumption  is questionable for penguin processes.
QCD factorization  \cite{BBNS} 
can be used to calculate
the non-factorizable diagrams.  We will use this framework to improve
the theoretical predictions for $B^{0,\pm}\to \phi\pi^{0,\pm}$. 
It is interesting to note that 
$B^{0,\pm}\to \phi\pi^{0,\pm}$ do not receive 
annihilation contribution,  while  $B^0 \to \phi\phi$ is a pure
penguin annihilation process. To the best of our knowledge, there 
is no realistic theoretical study of  $B^0 \to \phi\phi$.  
With respect to the topology of 
non-factorizable penguin diagrams for charmless $B$ decays, the decay 
$B^0\to \phi\phi$ is of interest. 
It can give us insight into the strength
of the annihilation topology in nonleptonic charmless $B$ decays which
is
still in dispute  \cite{BBNS, lisanda}. 
Experimentally, $B^0\to \phi\phi$   is relatively easy 
to identify.
We find ${\cal B}(B^0 \to\phi\phi)=2.1^{+1.6}_{-0.3}\times 10^{-9}$
and     
${\cal B}(B^\pm \to\phi\pi^{\pm} )=2.0^{+0.3}_{-0.1}\times 10^{-8}$ in
the SM. 
The smallness of the SM  predictions for these decays makes them
sensitive 
probes for flavor physics beyond the SM. We use the recent Babar upper
limits
${\cal B}(B^{0,\pm} \to\phi\pi^{0,\pm} )<1.6\times 10^{-6}$,
at $90\%$ CL  \cite{babar}, to obtain limits on the relevant 
R Parity Violating (RPV) couplings. 
We then use these limits to deduce  
the maximal possible enhancement  of ${\cal B}(B^0 \to\phi\phi)$
in RPV supersymmetry. Note that currently  ${\cal B}(B^0 \to\phi\phi)
<1.2\times 10^{-5}$ at $90\%$ CL  \cite{cleoomega}.
    
The experimental upper limits could be 
improved in BaBar and Belle. Measurement of any of these decays 
with ${\cal B}\gsim 10^{-7}$ 
will serve as an evidence for new physics.

In the SM, the relevant QCD corrected Hamiltonian is
\begin{equation}
\label{heff}
{\cal H}_{eff}
=-\frac{4G_{F}}{\sqrt{2}}
V_{tb} V_{td}^*
\sum_{i=3}^{10}
C_{i}O_{i}~.
\end{equation}
The operators in  ${\cal H}_{eff}$ relevant for $b \to d s \bar{s}$
are given in  \cite{buras}, where
at the scale $\mu=m_b$, $C_3=  0.014$, $C_4= -0.035$,
$C_5=  0.009$, $C_6= -0.041$,
$C_7= -0.002/137$,
$C_8=  0.054/137$,
$C_9= -1.292/137$,
$C_{10}= 0.262/137$.
Using the effective Hamiltonian and naive factorization
\begin{eqnarray}
A(B^- \to\phi\pi^{-})&=&-\frac{G_{F}}{\sqrt{2}}V_{tb} V_{td}^*
\left[\left( a_3 +a_5 \right)   
      -\frac{1}{2}\left( a_7 + a_9 \right)
    \right]\nonumber \\       
             &\times& f_{\phi}m_{\phi} F^{B\to\pi}_{+}(m^2_{\phi})
                \epsilon^{\phi}_{L}\cdot(p_{B}+p_{\pi}),
\end{eqnarray}
and $A(B^0 \to\phi\pi^{0})=\frac{1}{\sqrt{2}} A(B^- \to\phi\pi^{-})$
with $a_i\equiv C_{i}+C_{i+1}/N_c$.
The contributions of strong penguin operators arising from the
evolution from  $\mu=M_W$ to $\mu=m_b$ is very small due
to the cancellations between them: $C_{3}(m_b )\simeq -C_{4}(m_b )/3$
and 
$C_{5}(m_b )\simeq -C_{6}(m_b )/3$. Obviously  the amplitude is
dominated by
electroweak 
penguin. Using $f_{\phi}=254$ MeV,  $|V_{td}|=0.008$, $N_C =3$, and
the form factor
 $F^{B\to\pi}_{+}(0)=0.28$  \cite{ruckl, qcdsr, ukqcd, lattice}, 
we get ${\cal B}(B^\pm \to\phi\pi^{\pm} )
=2{\cal B}(B^0 \to\phi\pi^{0} )=2.9\times 10^{-9}$.

In the above calculations, non-factorizable contributions are neglected.
However, this neglect is questionable for penguin dominated 
$B\to\phi \pi$. The leading non-factorizable diagrams in Fig.1 should be
 taken into account.
 To this end, we employ the QCD factorization framework  \cite{BBNS}, 
which incorporates important theoretical aspects of QCD like color
transparency, heavy quark limit and hard-scattering, and allows us 
to calculate non-factorizable contributions systematically. 
In this framework, non-factorizable contributions to $B^- \to \pi^-
\phi$
can be obtained by calculating the diagrams in Fig.1.   
\begin{figure}[htbp] 
\scalebox{0.6}{\epsfig{file=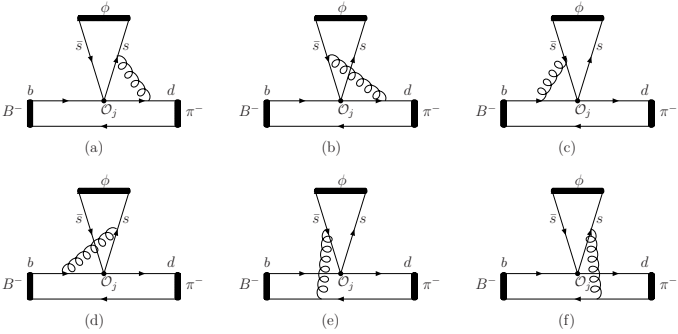}}
\caption{\label{fig1}Non-factorizable diagrams for $B^- \to
\phi\pi^{-}$.}
\end{figure}
To leading twist and leading power, the amplitude for $B^- \to\pi\phi$
is
\begin{eqnarray}
A'(B^- \to\phi\pi^{-})=-\frac{G_{F}}{\sqrt{2}}V_{tb} V_{td}^*
f_{\phi}m_{\phi} F^{B\to\pi}_{0}(m^2_{\phi})
                2\epsilon^{\phi}_{L}\cdot p_{B}
\nonumber \\
\times\biggl\{  
     \left[ 
         ( a_3 + a_5 )-\frac{1}{2}\left( a_7 + a_9 \right) 
      \right] 
+\frac{\alpha_{s}(\mu) }{4\pi} \frac{C_F}{N_c}
\biggr. \nonumber \\
 \times\biggl.
 \left[
            (C_4 -\frac{1}{2}C_{10}) F_{\phi}
              +(C_6 -\frac{1}{2}C_8 )(-F_{\phi}-12)   
                          \right]
\biggr\},~~ \label{nonfac}
\end{eqnarray}
and $A'(B^0 \to\phi\pi^{0})= \frac{1}{\sqrt{2}}A'(B^-
\to\phi\pi^{-})$.
The $\alpha_{s}$ term is the non-factorizable 
contribution with
\begin{eqnarray}
F_{\phi}=-12 \ln\frac{\mu}{m_b}-18+ V +
S,~~~~~~~~~~~~~~~~~~~~~~~~~~~~~~~\\
V=\int^1_0 du \Phi_{\phi}(u)\left( 3 \frac{1-2u}{1-u} \ln u
-3i\pi\right),
~~~~~~~~~~~~~~~~~\\
S=\frac{4 \pi^2}{N_c} \frac{f_{\pi}f_{B}}{M^2_B F_{+}^{B\to\pi}(0)}
\int^1_{0}d\xi dudv\frac{\Phi^{B}_{+}(\xi) }{\xi}
\frac{\Phi_{\phi}(u)}{u}
\frac{\Phi_{\pi}(v)}{v},~  
\end{eqnarray}
where $\xi=l_{+}/M_B$ is the momentum fraction carried by the spectator
quark in the $B$ meson. 
The $\Phi$'s are the leading twist 
light-cone distribution amplitudes of $\pi$, $\phi$ and $B$ mesons.
They describe the long-distance QCD dynamics of the matrix elements of
quarks 
and mesons, which is factorized out from the perturbative short-distance
interactions in the hard scattering  kernels.
These distribution amplitudes can be found in 
 \cite{grozin,schbda,pball}.   In our calculation,
we use the model proposed in \cite{schbda}
\begin{eqnarray}
\Phi^{B}_{+}(l_{+})&=&\sqrt{\frac{2}{\pi\lambda^2}}
\frac{l^2_+}{\lambda^2}
\exp\left[- \frac{l^2_+}{2\lambda^2}\right],\\
\Phi^{B}_{-}(l_{+})&=&\sqrt{\frac{2}{\pi\lambda^2}}
\exp\left[-\frac{l^2_+}{2\lambda^2}\right],
\end{eqnarray}
where $\lambda$ is the momentum scale of the 
light degrees of freedom in the $B$ and taken to  $350$ MeV. To show
model 
dependence of our prediction, we vary  $\lambda$ from 150 MeV to 550
MeV.  
we get 
${\cal B}(B^\pm \to\phi\pi^{\pm} )=2{\cal B}(B^0 \to\phi\pi^{0} )=
2.0^{+0.3}_{-0.1}\times 10^{-8}$.\ From 
Eq.\ref{nonfac}, we see that non-factorization is dominated by 
strong penguin due to the absence of $C_9$.  
We also note that non-factorizable contributions dominate these decays
and there is no isospin symmetry breaking because annihilation
contributions are absent.   

The decay $\bar{B}^0 \to \phi\phi$ is also of interest to study.
Firstly
it is a pure penguin process. Secondly  it is a pure annihilation and
thirdly its experimental signature is very clean.
By naive factorization,  the amplitude for this decay mode is 
\begin{eqnarray}   
A(\bar{B}^0 \to\phi\phi)=-4\frac{G_{F}}{\sqrt{2}}V_{tb}V_{td}^* 
 \left[ \left( a_3 -\frac{1}{2}a_9 \right) 
  \right. \nonumber\\   
~~~~~~~~\times
\langle \phi\phi|\bar{s}\gamma_{\mu}Ls|0\rangle
     \langle 0|\bar{d}\gamma^{\mu}Lb|\bar{B}^0 \rangle
 \nonumber\\
\left. +\left( a_5 -\frac{1}{2}a_7 \right) 
     \langle \phi\phi|\bar{s}\gamma_{\mu}Rs|0\rangle
\langle 0|\bar{d}\gamma^{\mu}Lb|\bar{B}^0 \rangle
\right] \nonumber \\
=-i4\frac{G_{F}}{\sqrt{2}}V_{tb} V_{td}^*  f_{B} p_{B}^{\mu}
\left[ \left( a_3 -\frac{1}{2}a_9 \right) 
     \langle \phi\phi|\bar{s}\gamma_{\mu}Ls|0\rangle
\right. \nonumber \\
\left. + \left( a_5 -\frac{1}{2}a_7 \right) 
     \langle \phi\phi|\bar{s}\gamma_{\mu}Rs|0\rangle
\right].
\end{eqnarray}
This amplitude vanishes for $m_s \to 0$. 
The $\alpha_s$ order matrix  
$\langle \phi\phi|\bar{s}\spur{p_B}(1-\gamma_5 )s|0\rangle$ also
vanishes 
due to the cancellation between the amplitudes of Fig.2.(c) and (d). 
\begin{figure*} 
\scalebox{1}{\epsfig{file=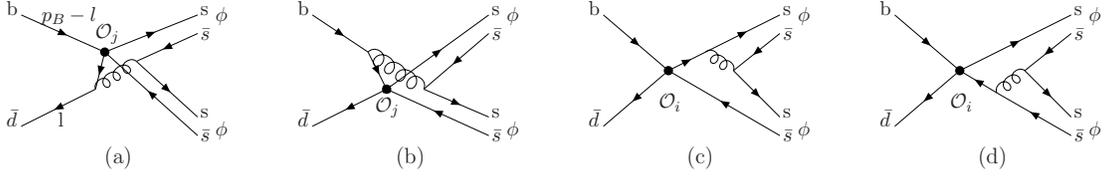}}
\caption{\label{fig2}(a) and (b) are non-factorizable diagrams 
for $\bar{B^0}\to \phi\phi$ decays. (c) and (d) are factorizable
diagrams
at $\alpha_s$ order.}
\end{figure*}
Non-factorizable contributions can be obtained by calculating the
amplitudes
of Fig.2.(a) and Fig.2.(b). They are
\begin{widetext}
\begin{eqnarray}
A(\bar{B}^0 \to\phi\phi)=\frac{G_{F}}{\sqrt{2}}V_{tb} V_{td}^*  
f_{B} f^2_{\phi} \pi \alpha_{s}(\mu_h )\frac{C_F }{N^2_C}
 \int^{\infty}_0 dl_{+} \int^1_0 dx  \int^1_0 dy
\Phi_{\phi}(x)\Phi_{\phi}(y)
\biggl\{ 
      \left( C_4 -\frac{C_{10}}{2}\right) 
        \left[
        \left( \xi -y \right)\Phi^{B}_{-}(l_{+})\frac{M_B^4}{D_b
k^2_g }
        \right.
\biggr. \nonumber\\        
\biggl. \left.
+ \left( x \Phi^{B}_{+}(l_{+})+ \xi\Phi^{B}_{-}(l_{+})\right) 
          \frac{M_B^4}{D_d k^2_g }
        \right]
\left( C_6 -\frac{C_{8}}{2}\right) 
      \left[
         y \Phi^{B}_{-}(l_{+})\frac{M_B^4}{D_d k^2_g }
       +\left[ \left( \xi -y \right)\Phi^{B}_{+}(l_{+})
               +\xi \Phi^{B}_{-}(l_{+})
         \right]\frac{M_B^4}{D_b k^2_g }
      \right]
    \biggr\},~~\label{aphi}   
  \end{eqnarray}
\end{widetext}
where we set $\mu_h$ to be the average virtuality of the time-like
gluon,
$\mu_h =M_{B}/2$. 
In Eq.\ref{aphi},
$k^2_g$ and $D_{b,d}$ are the virtualities of gluon, $b$ and $\bar{d}$ 
quark propagators, respectively. As in  \cite{example},
we meet end point divergence when $l_{+}=0$. 
Instead of a cut-off treatment \cite{example}, we 
use an effective gluon propagator  \cite{cornwall}
\begin{equation} 
\frac{1}{k^2}\Rightarrow \frac{1}{k^2 +M^2_{g}(k^2) },~~~~
M^2_{g}(k^2)=m^2_g \left[ \frac{\ln\left(\frac{k^2 +4
m^2_g}{\Lambda^2}\right)}
                            {\ln\left(\frac{4 m^2_g}{\Lambda^2}\right)}
                   \right]^{-\frac{12}{11}}.
\end{equation}
Typically $m_g =500\pm 200$ MeV, $\Lambda=\Lambda_{\rm {QCD}}=300$ MeV. 
Our use of this gluon propagator instead of imposing
a cut-off, is supported by lattice  \cite{glat}, and field 
theoretical solutions \cite{alkofer} which indicate that the
gluon propagator is not divergent as fast as $\frac{1}{k^2}$. 
Finally  we get ${\cal B}(B^0 \to\phi\phi)=2.1^{+1.6}_{-0.3}\times
10^{-9}$. 

Potentially, this decay may be enhanced by rescattering
$B\to\eta^{(\prime)}\eta^{(\prime)}
\to \phi\phi$ or by  $\omega-\phi$ through the channel $B^0 \to
\omega\phi\to\phi\phi$.
However, $\eta'$ and $\eta$ 
contributions are almost completely canceled\cite{petrov}, and  
$\phi$ is nearly a pure $\bar{s}s$ state, so the mixing mechanism
is also negligible. Furthermore in the language of QCD factorization
framework,
such kinds of soft final state interactions are subleading and
suppressed by power of 
${\cal O}(\Lambda_{QCD}/m_b)$\cite{neubert}, although it is hard to be
calculated 
reliably. Lastly, strong interaction annihilation
is negligible  since at least two gluons should be exchanged.
Thus, any unexpected large branching ratio observed, will 
indicate new physics.  

As an example for new physics, 
we will discuss the effects of the trilinear
$\lambda'$ and $\lambda''$ terms in the 
RPV  superpotential $W_{\spur{R}}$ \cite{MSSM, farrar, Rgroup} 
on the process $b\to ds\bar{s}$. We are therefore interested in
\begin{equation}
W_{\spur{R}}=\varepsilon^{ab}\delta^{\alpha\beta}\lambda^{'}_{ijk}L_{ia} 
Q_{jb\alpha}D^{c}_{k\beta} 
+\frac{1}{2}\varepsilon^{\alpha\beta\gamma}  
\lambda''_{i[jk]} U^{c}_{i\alpha} D^{c}_{j\beta} D^{c}_{k\gamma},   
\end{equation}
where $a,b$ are $SU(2)$ indices, $i,j,k$ are generation indices,
$\alpha,\beta,\gamma$ 
are $SU(3)$ color indices and $c$ denotes charge conjugation. 
The $L~(Q)$ are the lepton (quark) $SU(2)$ doublet
superfields, and
$U~(D)$ are the up- (down-) quark $SU(2)$ singlet
superfields.
Then we have 
\begin{eqnarray} 
{\cal L}_{\spur{R}}&=&\frac{1}{2} \varepsilon^{\alpha\beta\gamma} \left[ 
\lambda''_{ijk} \tilde{u}_{Ri\alpha} 
\left(\bar{d^c}_{j\beta} R d_{k\gamma}-\{j\leftrightarrow k\}\right)
\right] \nonumber\\
 &+&\lambda'_{ijk} \tilde{\nu}_{Li}\bar{d}_{k} L d_{j} 
+h.c.
\end{eqnarray}
\ From ${\cal L}_{\spur{R}}$, we get the effective Hamiltonian for $b\to
d s\bar{s}$
\begin{eqnarray}
{\cal H}_{\spur{R}}&=&
 -\frac{2}{m^2_{\tilde{u}_i} } \eta^{-4/\beta_0 }
\lambda''_{i23} \lambda''^{*}_{i12}
 \left[
       \left( \bar{s}_{\beta}\gamma_{\mu} R s_{\beta} \right) 
       \left( \bar{d}_{\gamma}\gamma^{\mu} R b_{\gamma} \right)
\right. \nonumber\\ 
&&\left.~~~~~~~~~~~~~~~~~~
 - \left( \bar{s}_{\beta}\gamma_{\mu} R s_{\gamma} \right) 
       \left( \bar{d}_{\gamma}\gamma^{\mu} R b_{\beta} \right)
\right] \nonumber\\
&&-\frac{1}{2m^2_{\tilde{\nu}_i}}\eta^{-8/\beta_0 }
 \left[ 
   \lambda'_{i31}\lambda'^{*}_{i22}(\bar{s}_{\alpha} \gamma_{\mu} L
   b_{\beta}) (\bar{d}_{\beta} \gamma^{\mu} R s_{\alpha})
\right. \nonumber\\
&&~~~~~~~~
+\lambda'^{*}_{i13}\lambda'_{i22}(\bar{s}_{\alpha} \gamma_{\mu} R
b_{\beta}) (\bar{d}_{\beta} \gamma^{\mu} L s_{\alpha}) 
   \nonumber\\
&&~~~~~~~~~
 +\lambda'_{i32}\lambda'^{*}_{i12}(\bar{d}_{\alpha} \gamma_{\mu} L
b_{\beta}) (\bar{s}_{\beta} \gamma^{\mu} R s_{\alpha})
\nonumber \\
&&\left.
~~~~~~~
+\lambda'^{*}_{i23}\lambda'_{i21}(\bar{d}_{\alpha} \gamma_{\mu} R
b_{\beta}) 
(\bar{s}_{\beta} \gamma^{\mu} L s_{\alpha})\right], 
\end{eqnarray}
where $\eta=\frac{\alpha_{s}(m_{\tilde{f}_i})}{\alpha_{s}(m_b )}$ and
$\beta_0 =11-\frac{2}{3}n_f $. The coefficients $\eta^{-4/\beta_0 }$
and $\eta^{-8/\beta_0 }$ are
due to running from the sfermion mass 
scale $m_{\tilde{f}_i}$ (100 GeV assumed) down to the
$m_b$ scale.

We can now write down the contributions of ${\cal H}_{\spur{R}}$ to
$B^- \to \phi\pi^{-}$ and $B^0 \to \phi\phi $ decays
\begin{widetext}
\begin{eqnarray}
 A^{\spur{R}}(B^- \to \phi\pi^{-})&=& 
-\frac{1}{8 m^2_{\tilde{\nu}_i}} \left( \lambda'_{i21}
\lambda'^{*}_{i23}
+\lambda'^{*}_{i12} \lambda'_{i32} \right) \eta^{-8/\beta_0 }
f_{\phi} F{B\to\pi}(m^2_{\phi}) M^2_B 
\left[ \frac{1}{N_c} 
      +\frac{\alpha_s }{4\pi}\frac{C_F}{N_c}\left(- F_{\phi}-12\right) 
\right]\nonumber\\                                     
&&-\frac{1}{2 m^2_{\tilde{u}_i}} \lambda''_{i23}
\lambda''^{*}_{i12}\eta^{-4/\beta_0 } 
f_{\phi} F^{B\to\pi}(m^2_{\phi}) M^2_B 
\left( \frac{2}{3}-\frac{\alpha_s }{4\pi} \frac{C_F}{N_c} F_{\phi}
\right),
\\
A^{\spur{R}}(B^0\to\phi\phi)&=&
-\frac{1}{2 m^2_{\tilde{u_i}}}  \lambda''_{i23} \lambda''^{*}_{i12}
\eta^{-4/\beta_0 }
f_{B} f^2_{\phi} \pi \alpha_{s}(\mu_h )\frac{C_F }{N^2_C}
 \int^{\infty}_0 dl_{+} \int^1_0 dx  \int^1_0 dy
\Phi_{\phi}(x)\Phi_{\phi}(y)\nonumber \\
&& \left[
          \left( x \Phi^{B}_{+}(l_{+})+
\xi_{+}\Phi^{B}_{-}(l_{+})\right) 
          \frac{M_B^4}{D_d k^2_g }
        +\left( \xi_{+}-y \right)\Phi^{B}_{-}(l_{+})\frac{M_B^4}{D_b
k^2_g }
        \right] \nonumber \\
&-&\frac{1}{8 m^2_{\tilde{\nu}_i}} 
\left( \lambda'_{i32} \lambda'^{*}_{i12} 
+\lambda'^{*}_{i23} \lambda'_{i21} 
\right)
\eta^{-8/\beta_0 }
f_{B} f^2_{\phi} \pi \alpha_{s}(\mu_h )\frac{C_F }{N^2_C}
 \int^{\infty}_0 dl_{+} \int^1_0 dx  \int^1_0 dy
\Phi_{\phi}(x)\Phi_{\phi}(y)\nonumber \\
&& \left[
          y \Phi^{B}_{-}(l_{+}) \frac{M_B^4}{D_d k^2_g }
        +\left( (\xi_{+}-y) \Phi^{B}_{+}(l_{+}) 
                +\xi_{+} \Phi^{B}_{-}(l_{+})
         \right) 
\frac{M_B^4}{D_b k^2_g }
        \right]. 
\end{eqnarray}
\end{widetext}

In the numerical results, we assume that only one sfermion
contributes at a time and that they all have a mass of 100 GeV. 
The uncertainties of the theoretical predictions, due mainly to the B meson 
distribution function, are displayed as thickness of curves in Fig.3 
  
Our results for the RPV
contributions to $B \to \phi\pi$  
are summarized in Fig.3.\ From the BaBar upper limit  \cite{babar} 
${\cal B}(B^{0,\pm} \to\phi\pi^{0,\pm} )<1.6\times 10^{-6}$, we obtain 
the following constraints ($90\% CL$) 
\begin{eqnarray}
| \lambda^{''}_{i23}\lambda^{''*}_{i21}|&<&6\times10^{-5}
\left(\frac{m_{\tilde{u}_{Ri}}}{100}\right)^2 ~,\\
| \lambda^{'}_{i32}\lambda^{'*}_{i12}|&<&4\times10^{-4}
\left(\frac{m_{\tilde{\nu}_{Li}}}{100}\right)^2 ~,\\  
| \lambda^{'}_{i21}\lambda^{'*}_{i23}|&<&4\times10^{-4}
\left(\frac{m_{\tilde{\nu}_{Li}}}{100}\right)^2 ~.
\end{eqnarray}

We note that our constraints on $\lambda^{''}_{i23}\lambda^{''*}_{i21}$
are more than one order of magnitude stronger  than the limits 
obtained recently  \cite{he}.
For 
$\lambda^{'}_{i32}\lambda^{'*}_{i12}$ and 
$\lambda^{'}_{i21}\lambda^{'*}_{i23}$, our bounds are comparable with
the present upper limits  \cite{he, Allanach:1999ic}. 
Within the available upper bounds for these couplings,
$\lambda^{''}_{i23}\lambda^{''*}_{i21}$ RPV terms could enhance
${\cal B}(B^0 \to \phi\phi)$ to  $10^{-8}$, while
$\lambda'_{i23} \lambda'^{*}_{i21}$ and $\lambda'_{i32}
\lambda'^{*}_{i12}$ 
RPV terms could enhance ${\cal B}(B^0 \to \phi\phi)$ to 
$10^{-7}$ which may be measurable at Belle and Babar. 

In summary, we have studied the pure penguin processes
$B^{\pm,0}\to\phi\pi^{\pm,0}$ and 
$B^{0}\to \phi\phi$ by using QCD factorization for the hadronic
dynamics. We estimate that in the SM
${\cal B}(B^- \to \phi\pi^- )=2.0^{+0.3}_{-0.1}\times10^{-8}$ and 
${\cal B}(B^0 \to \phi\phi )=2.1^{+1.6}_{-0.3}\times10^{-9}$. The 
smallness of these decays in the SM makes them sensitive probes of 
flavor physics beyond the SM. Using the BaBar result  
${\cal B}(B^- \to \phi\pi^- )<1.6\times10^{-6}$, we have 
obtained new bounds on some products of RPV  coupling constants. 
\begin{figure}[htbp] 
\scalebox{0.5}{\epsfig{file=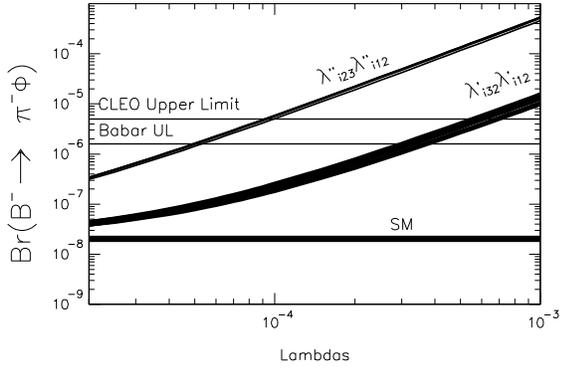}}
\caption{\label{fig3}The branching ratio of $B^- \to\phi\pi^-$ as a
function of
the RPV couplings $|\lambda''_{i23} \lambda''^{*}_{i12}|
$(upper curve),
$|\lambda'_{i23} \lambda'_{i21}|$ and $|\lambda'_{i32} \lambda'_{i12}|$
(lower curve) respectively. 
The thickness of curves represent our theoretical uncertainties. 
The horizontal lines are the upper limits and the SM prediction as 
labeled  respectively. The thicknesses of the curves and the line
labelled as SM 
are theoretical uncertainties.}
\end{figure}
In the case of $\lambda''_{i23} \lambda''^{*}_{i12}$, our limits are 
better than previous bounds. 
Given the available bounds on   
$\lambda^{''}_{i23}\lambda^{''*}_{i21}$, 
$\lambda'_{i23} \lambda'^{*}_{i21}$ and $\lambda'_{i32}
\lambda'^{*}_{i12}$,
the decay  $B^0 \to \phi\phi$ could be enhanced to
$10^{-8}\sim10^{-7}$.  
Due to the clear signatures of $\phi$ and $\pi^\pm$, the experimental 
sensitivity of these decay modes is high. Babar and Belle could
reach very low upper limits on these decays if not measured. 
Searches for these decays are strongly urged.


We thank Y. Grossman, M. Gronau and S. Roy  for helpful discussions.
This work is supported by the US-Israel Binational Science
Foundation and the Israel Science Foundation. 
%
%
%


\begin{thebibliography}{99}
\bibitem{pdg2000} Particle Data Group, E. Groom  $et\,al$., Eur. Phys.
J. {\bf C15}, 1 (2000).
\bibitem{BBNS} M. Beneke $et\, al.$, 
Phys. Rev. Lett. {\bf 83}, 1914 (1999); Nucl. Phys. {\bf B579},
313 (2000). 
\bibitem{lisanda} Y.Y. Keum, H.N. Li and A.I. Sanda,  
Phys. Rev. {\bf D63}, 054008 (2001);  C.D. L\"u, K. Ukai and M.Z. Yang,
Phys. Rev. {\bf D63},074009 (2001).  
\bibitem{fleischer}R. Fleischer, Phys. Lett, {\bf B321}, 259 (1994).
\bibitem{DU} D.S. Du and L.B. Guo, Z. Phys. {\bf C75}, 9 (1997);
\bibitem{babar} BaBar Collaboration: D. Harrison, BaBar-talk-02/08, 
at Aspen Winter Conference, 2002.
\bibitem{cleoomega} CLEO Collaboration: T. Bergfeld  $et\,al.$, 
Phys. Rev. Lett. {\bf 81}, 272 (1998).
\bibitem{buras} For a review, see G. Buchalla, A.J. Buras and 
M.E. Lautenbacher, Rev. Mod. Phys. {\bf 68}, 1125 (1996).
\bibitem{ruckl} A. Khodjamirian $et\,al$., 
Phys. Rev. {\bf D62}, 114002 (2000).
\bibitem{qcdsr} E. Bagan, P. Ball and V. M. Braun, Phys. Lett. {\bf
B417}, 154 (1998);
P. Ball, JHEP {\bf 09}, 005 (1998).
\bibitem{ukqcd} UKQCD Collaboration: K.C. Bowler $et\,al$., Phys. Lett.
{\bf B486}, 111 (2000).
\bibitem{lattice} A. Abada, $et\,al$., Nucl. Phys. {\bf B416}, 675
(1994), ibid., {\bf B619}, 565(2001). C.R. Allton,
$et\,al$.,  Phys. Lett. {\bf B345}, 513 (1995).
\bibitem{grozin} 
A.G. Grozin and M. Neubert, Phys. Rev. {\bf D55}, 272 (1997);
M.Beneke and F. Feldmann, Nucl. Phys. {\bf B592}, 3(2001).
\bibitem{schbda} 
S. Descotes-Genon and C.T. Sachrajda, Nucl. Phys. {\bf B625}, 239 (2002).
\bibitem{pball}
P. Ball and V.M. Braun, Phys. Rev. {\bf  D58}, 094016;
P. Ball, JHEP {\bf 9901}, 010 (1999).
\bibitem{example}
M. Beneke $et\,al$., Nucl. Phys. {\bf B606}, 245(2001);
S.W. Bosch and G. Buchalla, Nucl. Phys. {\bf B621}, 459 (2002);
M. Beneke, T. Feldmann and D. Seidel, Nucl. Phys. {\bf B612}, 25 (2001).
\bibitem{cornwall} J.M. Cornwall, Phys. Rev. {\bf D26}, 1453 (1982);
J.M. Cornwall and J. Papavasiliou, Phys. Rev. {\bf D40}, 3474 (1989),
$ibid$, {\bf D44}, 1285 (1991). 
\bibitem{glat}
A.G. Williams $et\, al$., hep-ph/0107029 and references therein.
\bibitem{alkofer} R. Alkofer and L. von Smekal, Phys. Rept. {\bf353},
281 (2001), and references therein.
\bibitem{petrov} A.A. Petrov, hep-ph/9909312.
\bibitem{neubert}M. Neubert, hep-ph/0012204.
\bibitem{MSSM}For review see  H. Dreiner, 
in {\it Perspectives on Supersymmetry}, Ed. by G.L. Kane, World
Scientific.
\bibitem{farrar}G. Farrar and P. Fayet, Phys. Lett. {\bf B76},
575 (1978).
\bibitem{Rgroup} R. Barbier, $et\, al$., hep-ph/9810232.
\bibitem{he} D.K. Ghosh, $et\, al$., hep-ph/0111106.
\bibitem{Allanach:1999ic}
C. Allanach, A. Dedes and H.K. Dreiner,
Phys.\ Rev.\ D {\bf 60}, 075014 (1999).
\end{thebibliography}
\end{document}